\documentclass[12pt,preprint]{elsarticle}

\usepackage{amssymb,amsmath}
\usepackage{color}
\usepackage{physics}
\usepackage{centernot}
\usepackage{siunitx}
\usepackage{csquotes}

\journal{Journal of Magnetism and Magnetic Materials}

\begin{document}

\begin{frontmatter}



\title{Tuning properties of phase-separated magnetic fluid with temperature}


\author[LU]{Laura Nelsone}
\author[LU]{Guntars Kitenbergs}
\ead{guntars.kitenbergs@lu.lv}


\affiliation[LU]{organization={MMML lab, Department of Physics, University of Latvia},
            addressline = {Jelgavas~iela~3}, 
            city={Riga},
            postcode={LV-1004}, 
            country={Latvia}}

\begin{abstract}
Phase-separated magnetic fluids provide a very strong magnetic response ($\mu>25$) in a liquid state material. 
Even small fields can cause a notable material response, but this depends on its properties, which are often difficult to control.
Here, we investigate how temperature affects the properties of the system, where phase separation is induced by an increase in ionic strength.
Following the deformations of individual microscopic droplets, we can extract surface tension, magnetic permeability, and viscosity.
We find that the temperature increase does not affect the surface tension, while the magnetic permeability and viscosity increase.
Applying this knowledge to induce a shape instability for a drop only using temperature shows the potential of applicability of our findings.

\end{abstract}




\end{frontmatter}


\section{Introduction}
\label{sec:intro}

Magnetic fluids are made of magnetic particles dispersed in a carrier fluid. 
Today, they are used in various industrial applications, for example, brakes, dampers, and loudspeakers~\cite{scherer2005ferrofluids}. 
More recently, there has been an increase in biomedical use, including biosensors and disease treatment~\cite{zhang2019flexible}. 
Such a wide application variety is due to the unique combination of properties, the ability to react to an applied magnetic field, and the fluidity it retains even when it is exposed to a strong magnetic field \cite{zhang2019flexible}.
When two phases with different magnetic properties are combined, magnetic droplets can be observed.
Recently, a variety of interesting and promising applications have been proposed for them, including a model system for active matter \cite{stikutsdrops} or microreactors \cite{Rigonidroplets}, tissue mechanics detection \cite{tissuemechanics}, and even controllable robots \cite{PNASrobots}.
There are several approaches to obtain magnetic fluid droplets.
The simplest is to use two immiscible fluids, where one of the fluids is magnetic (for example, see \cite{droplargefield}).
A similar, but more peculiar way is to add magnetic particles to stabilize the interface of immiscible fluids, creating a magnetic Pickering emulsion \cite{pickering}.
An approach closer to biology is to use a liposome, which is filled with a magnetic fluid~\cite{Magnliposomes}.
An alternative way is to use the colloidal phase separation phenomenon \cite{phasesep}, which leads to two phases with different concentrations of magnetic particles.
One phase typically has a high particle concentration, whereas the other is more dilute, as the initial magnetic fluid.
Typically, the surface tension between the two phases is very low.
This can be achieved with a solution of two polymers and magnetic nanoparticles \cite{rigoni2020ferrofluidic}.
However, a more common method is to destabilize a charged colloidal magnetic fluid.
The resulting magnetic droplets are strongly magnetic and, due to the low surface tension, the droplets can deform into a large variety of shapes even under small magnetic fields \cite{dropPRL,erdmanis2017magnetic}, further widening the possible applications. 

To enable the variety of applications, it is of interest to adjust the properties of the material.
Among the factors that influence phase separation in magnetic fluids, the ionic strength (through the addition of NaCl) and the temperature can be considered equivalent \cite{langmuir_og}.
If adding salt to a magnetic fluid is easy when preparing the material, then the temperature can also be tuned during an experiment.
Here, we investigate this possibility by determining the parameters for the concentrated magnetic fluid phase, which forms droplets at different temperatures.
We approach this by analyzing droplet shape deformations during an elongation and relaxation experiment under small magnetic fields \cite{erdmanis2017magnetic} for many droplets across a range of temperatures.
It should be noted that droplet shape analysis has also been used to investigate other phase-separated systems \cite{atefi2014ultralow}, but without investigating the temperature dependence.

\section{Theory}
\label{sec:theory}
To experimentally determine the surface tension, magnetic permeability, and viscosity of magnetic droplets, they can be stretched out in a stepwise manner in an external magnetic field and allowed to relax back to a spherical shape from a stretched state at a zero external magnetic field \cite{erdmanis2017magnetic}.

To find surface tension and magnetic permeability, it is assumed that the magnetic permeability of the droplet is constant, the surrounding fluid is nonmagnetic and that all shape deformations can be described with an ellipsoid. 
In this theoretical model, it is assumed that an external magnetic field creates magnetic energy in the droplet which makes it stretch out in the direction of the magnetic field. 
At the same time, surface tension attempts to keep the shape with as little surface as possible. 
A stable shape is reached once the magnetic energy and the surface tension energy are in balance \cite{bacri1982instability}.
Experimentally, this can be reached by keeping a magnetic droplet in a constant magnetic field until the shape of the droplet reaches equilibrium. 

The magnetic droplet eccentricity \(e =\sqrt{1-(b/a)^2}\) in which $a$ is the semi-major axis of the ellipse and $b$ is the semi-minor axis has been determined to maintain the following relation with the external magnetic field applied (H) and the initial magnetic droplet radius $R$ \cite{bacri1982instability,cebersh2r}:

\begin{equation}
    \label{sigma_mu_eq}
    H^2R = \gamma \left[\frac{4\pi}{\mu-1} + N\right] ^2 \frac{1}{2\pi} \frac{\left(\frac{3-2e^2}{e^2} - \frac{(3-4e^2)\arcsin e}{e^3(1-e^2)^{1/2}} \right)} 
    {(1-e^2)^{2/3} \left( \frac{3-e^2}{e^5} \ln{\frac{1+e}{1-e}} - \frac{6}{e^4} \right)},
\end{equation}

\noindent in which \(\gamma \) - magnetic droplet surface tension coefficient, \(\mu\) - magnetic droplet's magnetic permeability and \textit{N} - demagnetization coefficient, which for an ellipse is:

\begin{equation}
    \label{N_value}
    N = \frac{4\pi(1-e^2)}{2e^3} \left( \ln{\frac{1+e}{1-e}} - 2e \right)
\end{equation}

A useful quantity to note here is the magnetic Bond number typically defined as $Bm=H^2R/\gamma$.
This allows us to compare measurements for droplets of different sizes.

\begin{figure}
    \centering
    \includegraphics[width=\textwidth]{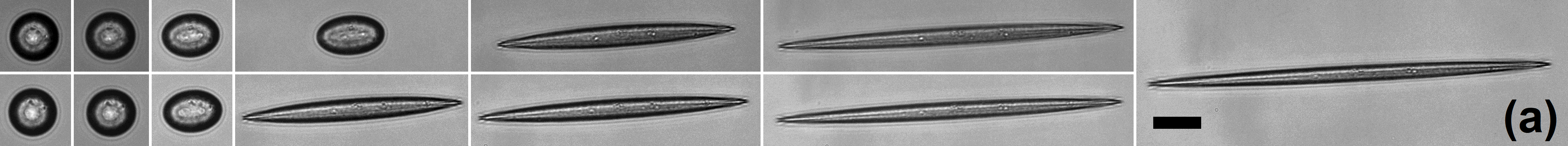}
    \includegraphics{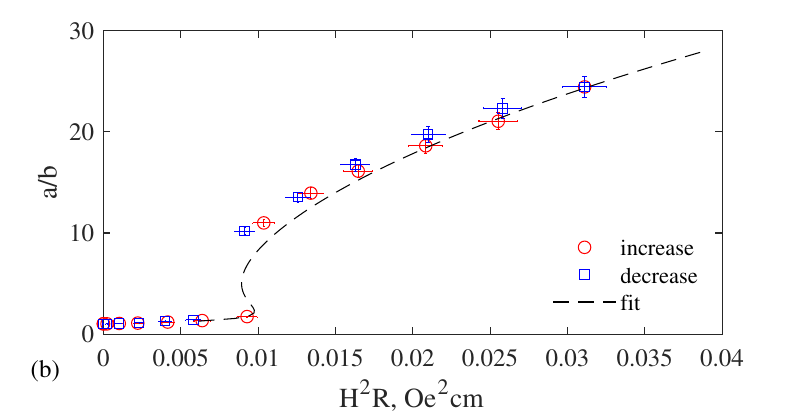}
    \caption{Example of experimental measurements indicating hysteresis, when $\mu\approx30$. Images in (a) show droplet shapes when field is increased in a stepwise manner (upper row from left to right) and decreased (lower row from right to left). Images are combined so that magnetic field values are equal. Scale bar is $20~\mu$m. Proportion of the corresponding droplet longer and shorter semi-axes $a/b$ is visualized in (b) both during a field increase (red circles) and decrease (blue squares). Black dashed line shows the best fit with \eqref{sigma_mu_eq}, giving $\mu=28.5$ and $\gamma=2.6\cdot10^{-3}$~dyn/cm.}
    \label{fig:hyst_explanation}
\end{figure}

It is a well-known fact that a hysteresis process can be observed for the deformation of a magnetic droplet during increase and decrease of an external magnetic field. 
The shape difference can be observed if the magnetic field is gradually increased before reaching the bifurcation point and after that point by gradually decreasing the magnetic field until the second bifurcation point is reached \cite{Cebera_bible}. 
There is a certain region in which a magnetic droplet can assume two stable shapes with different values of $a/b$ if specific conditions are met for the phase separated fluid. 
The theoretical model mentioned above also predicts this transition for \(\mu > 21\). 
This has been shown experimentally and numerically in several other works. 
As shown in later sections, also in this study, our magnetic droplets behave in this way, as shown in an example in Figure \ref{fig:hyst_explanation}.

In literature it has been shown that the ellipsoidal approximation works only when the droplet shapes do not show hysteresis, meaning that magnetic permeability $\mu < 21 $ or the magnetic droplet is not stretched above $a/b\approx4$~\cite{Dropsimulation}.
Once these conditions are overstepped, the magnetic droplet creates conical tips due to nonuniformity of the magnetic field inside the magnetic droplet \cite{misra2020magnetic}. 
This phenomenon has been known for a long time; however, not much literature attempts to describe the shape of the magnetic droplet after the hysteresis jump. 

To solve the problem of magnetic droplet shape once conical tips are created, Misra proposed a different way of describing the magnetic droplet shape \cite{misra2020magnetic}. 
Instead of assuming a homogeneous magnetic field inside the droplet, it is acknowledged that it is inhomogeneous; thus, the volume average of the magnetic field and magnetization inside the droplet are used to find a semi-analytical shape of the magnetic droplet. 
The resulting equation is as follows:

\begin{equation}
    \tilde{z} = \pm \frac{1}{1-cos\theta_c} \sqrt{-(1-cos\theta_c)^2\tilde{r}^2 - 2cos\theta_c(1-cos\theta_c)\tilde{r} + (1-cos\theta_c^2)} 
    \label{misra_eq}
\end{equation}

\noindent in which $\tilde{z}$ and $\tilde{r}$ are normalized values of values $a$ and $b$ accordingly. 
Both $\tilde{z}$ and $\tilde{r}$ correspond to coordinates that describe the length and width of the magnetic droplet at every point. 
$\theta_{c}$ is the half-cone angle of the magnetic droplet's cone. 
After finding $\theta_{c}$ in Eq.~(\ref{misra_eq}), the $a/b$ proportion can be calculated as 

\begin{equation}
    \frac{a}{b} = \sqrt{\frac{1+cos\theta_c}{1-cos\theta_c}}
    \label{misra_prop}
\end{equation}

The third property of magnetic fluid that can be found in experiments is viscosity \(\eta\).
It is related to the characteristic rate at which the magnetic droplet relaxes back to a spherical shape at zero external magnetic field from an elongated shape \cite{dikansky1990magnetic}. 
Quantitatively, it has been determined that the relaxation time of the magnetic droplet shape \(\tau\) is correlated with the ambient viscosity \(\eta_v\) and magnetic droplet's viscosity \(\eta\):

\begin{equation}
    \label{tau_eq}
    \tau = \frac{R(16\eta_v+19\eta)(3\eta_v+2\eta)}{40\gamma(\eta_v+\eta)}
\end{equation}

Droplet deformation can be characterized as $(a-b)/a$, while $\tau$ can be obtained from the exponent $e^{-t/\tau}$ when it is close to a spherical shape, as previously shown \cite{erdmanis2017magnetic}.
This method is very similar to the retraction analysis of a droplet deformed by shear flow \cite{guido}. 
It is worth mentioning that additional methods have been proposed \cite{mo,Yu}, which can be useful for a different set of physical parameters.

The surface tension, magnetic permeability, and viscosity values can be considered to be quasi-stable, meaning the values change over time.
However, in the span of a single experimental measurement, we consider these values to be constant if the temperature is kept constant. 

\section{Experiments}
\label{sec:setup}
\subsection{Materials and experimental setup}

The magnetic fluid used in this study was made in the PHENIX laboratory at CNRS and Sorbonne University, Paris, France, following the Massart method~\cite{MassartMF}. 
Nanoparticles were coated with citrate ions and stabilized at pH 7. 
The resulting colloid has maghemite particles with a particle volume fraction $\phi = (4.6-5.0)\%$, magnetic particle diameter $(7.0 \pm 0.3)$~nm and free citrate concentration $0.05$~mol/L.

Adding a 2.5 $\%v/v$ solution of 4 mol / L NaCl salt solution to the fluid destabilizes it and induces phase separation.  
6 $\mu$L of the phase separated magnetic fluid are placed in a rectangular glass capillary (\emph{Vitrocom}, $\approx25\times 0.1 \times 2$~mm$^3$) and both ends of the capillary are closed with a special sealant (\emph{Brand}) and covered with a UV glue \emph{Crystalbond}.
This ensures that the magnetic fluid does not evaporate from the capillary.

\begin{figure}
    \centering
    \includegraphics[width=.8\textwidth]{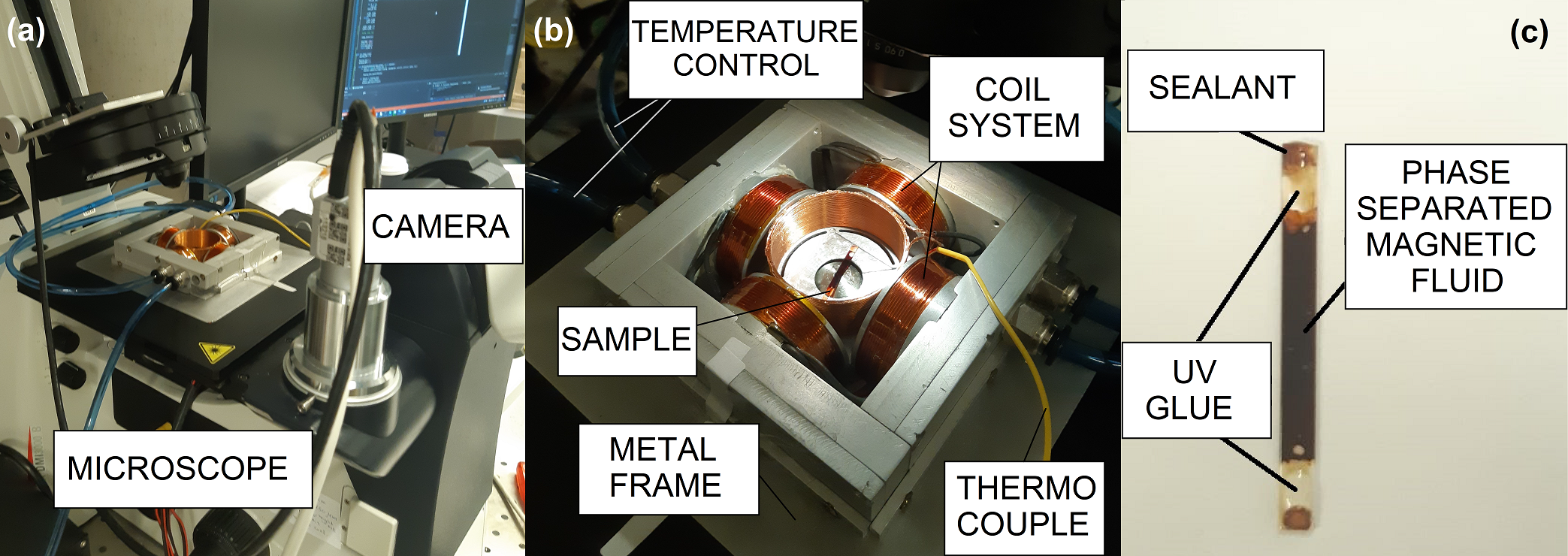}
    \caption{Experimental setup consists of a microscope and a camera (a), which is equipped with a temperature controlled coil system (b), where a sample in a sealed capillary is placed (c).}
    \label{fig:setup}
\end{figure}

Experimental system is introduced in Figure~\ref{fig:setup}.
Experiments are done on a \emph{Leica DMI3000 B} microscope, which is equipped with a camera \emph{Basler ac190-155um} and a $40\times$ objective in bright field mode. 
The magnetic field is created by an in-house built triaxial coil system, compatible with the microscope.
The intensity and direction of the magnetic field can be controlled in 3D with a \emph{Labview} program, which sends a voltage signal to 3 power supplies (\emph{KEPCO}), generating the corresponding current that runs through each pair of the coils and creating a homogeneous magnetic field across the field of view of the microscope.

Temperature control for both cooling and heating of the system is achieved by a thermostat (\emph{Huber Ministat 230}), which is connected to the metallic parts of the coil system, including the sample holder. 
The thermostat has a feedback loop with a thermocouple placed in contact with the metal frame. 
As the samples used in the experiment are small and were placed on the metal frame, the sample temperature was assumed to be the same as the metal frame temperature.
The temperature is registered with a thermocouple, which is connected to the sample holder. 

\begin{figure}
    \centering
    \includegraphics[width=.7\textwidth]{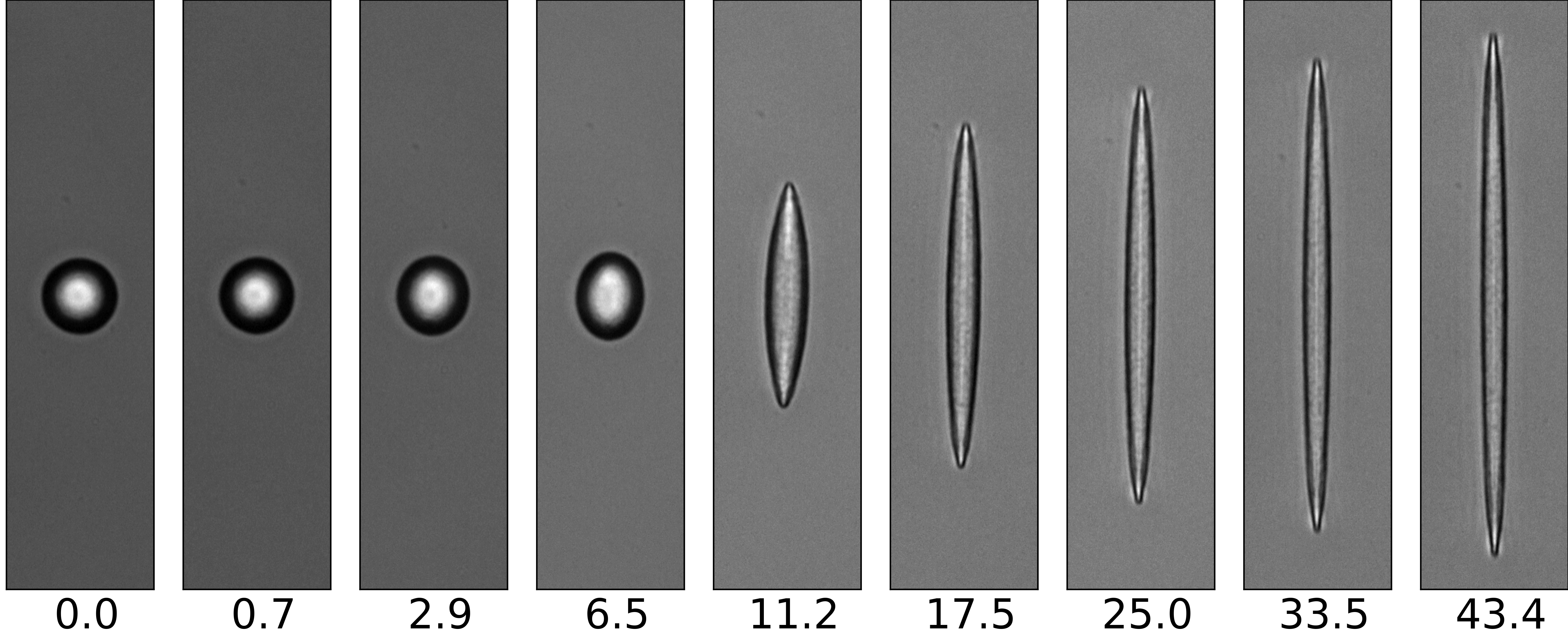}
    \includegraphics[width=.7\textwidth]{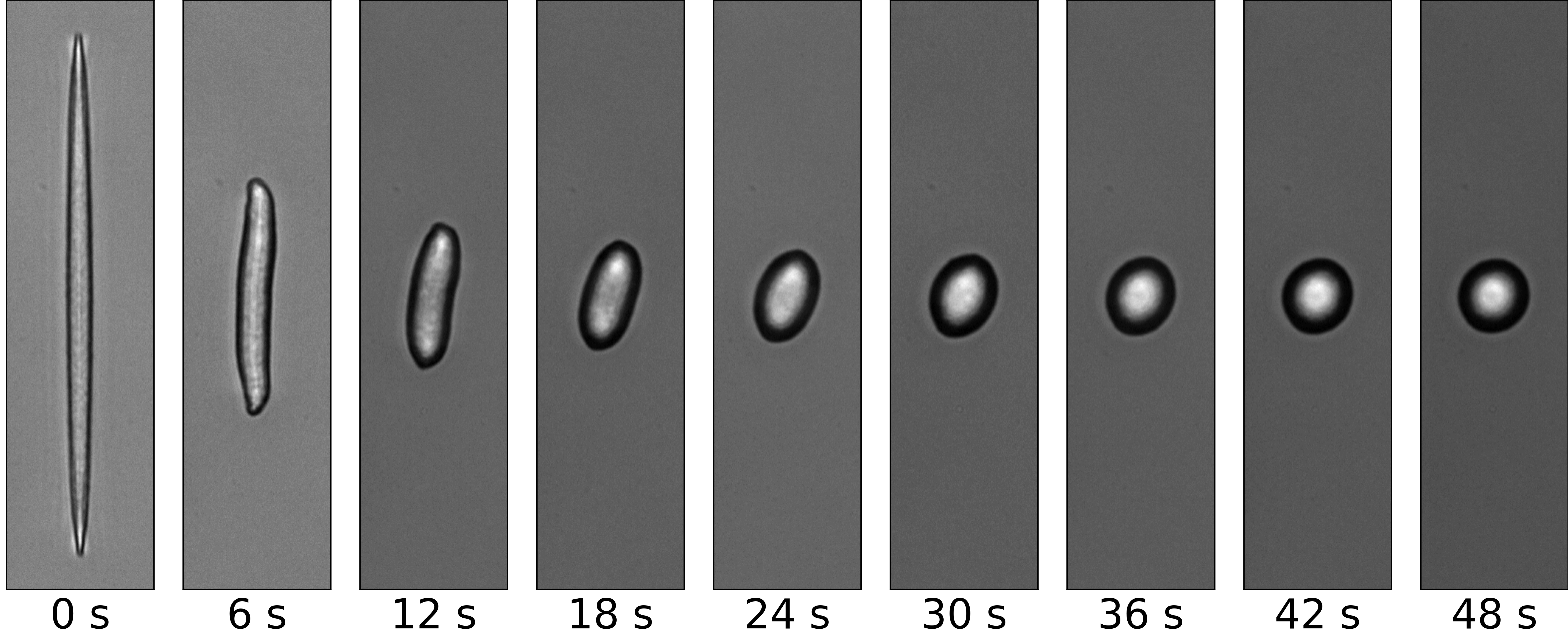}
    \caption{An example of elongation and relaxation of a magnetic droplet. Upper row shows step-wise stretching, where values represent $H^2R$ [mOe$^2\cdot$cm]. Lower row shows the time series of a magnetic droplet contracting to relaxed state at $H=0$.}
    \label{fig:droplet_stretching}
\end{figure}

For experiments, a single or several nearby magnetic droplets are placed in the field of view. 
Then a small and static magnetic field value is set and the magnetic droplet is allowed to reach an equilibrium shape, as is observable with the microscope camera.
The process is repeated in a stepwise manner.
An example of droplet deformations is shown in Fig.~\ref{fig:droplet_stretching} (upper row).
Once the droplet has reached a sufficiently large elongation, the magnetic field is either lowered in the same stepwise manner or suddenly turned off.
For the second case, an example of droplet relaxation is shown in Fig.~\ref{fig:droplet_stretching} (lower row).
The camera captures droplet shape deformations all the time.
As the process is slow, a frame rate of 2.5 frames per second is used.
Simultaneously, magnetic field and temperature data are registered.
The process is repeated for multiple droplets and at various temperatures.
After the change in temperature, the sample is left to equilibrate, typically for several hours.
Also, the light intensity of the bright-field imaging mode is kept reasonable, as it has the potential to heat up and change the characteristic properties of droplets. 

\subsection{Image and data processing}

To extract physical data from recorded images and measurement series, image and data processing and fitting steps are performed.
This is done by a self-made code in python.

First, several image processing steps are taken to find the outline of the droplet using \textit{scikit-image} library.
This includes image binarization with Otsu thresholding (\textit{threshold\_otsu}) and morphology operations (\textit{closing, dilation, erosion}).

\begin{figure}
    \centering
    \includegraphics[width=.7\textwidth]{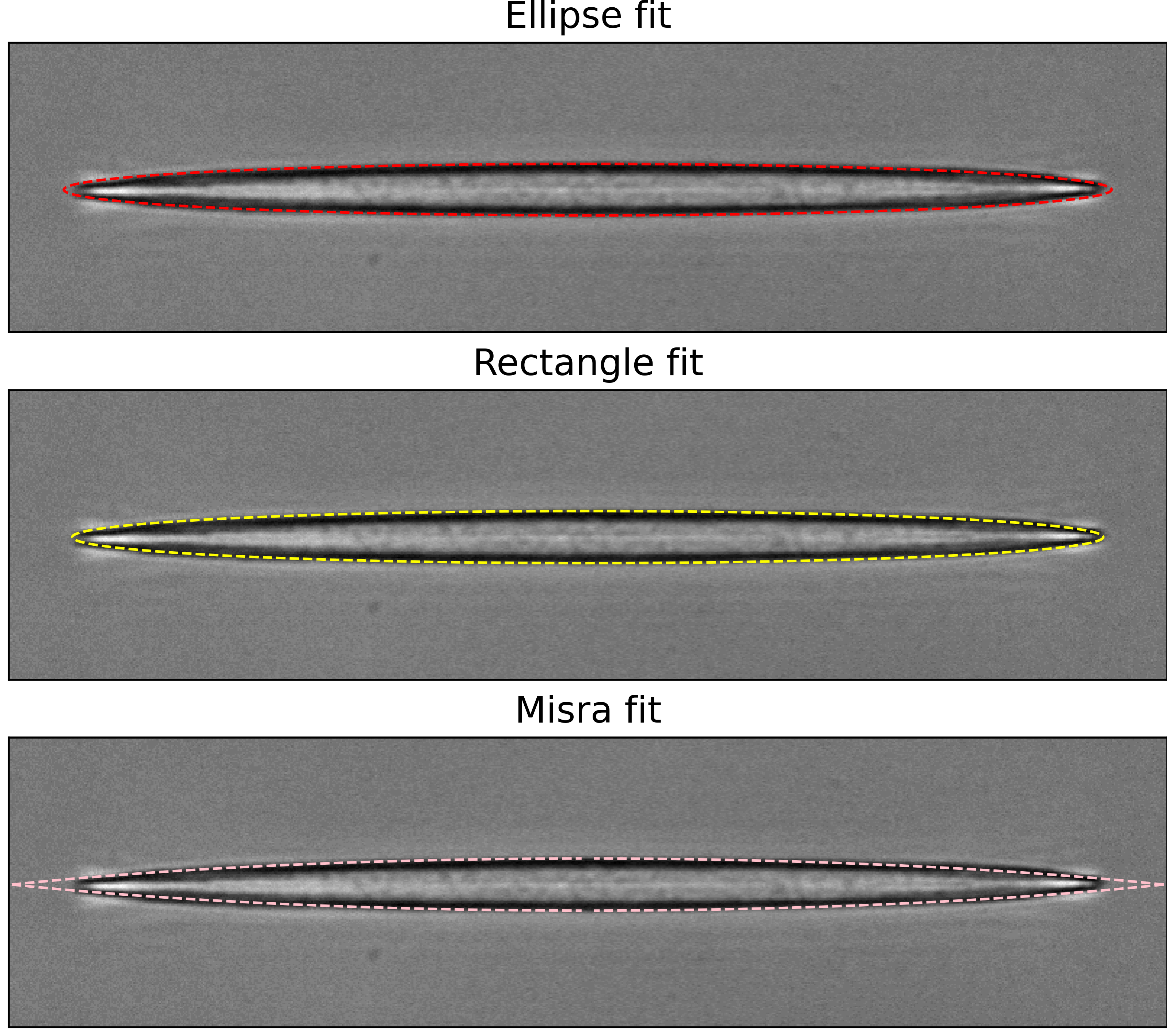}
    \caption{An example of different fits on an elongated magnetic droplet.}
    \label{fig:Misra_fit}
\end{figure}

Once the droplet outline data are available, an approximation can be used to describe it. 
An investigation was carried out on which shape - ellipsoidal (\textit{fitEllipse} from \textit{openCV}), rectangular (\textit{minAreaRect} from \textit{openCV}) or shape proposed by Misra (for simplicity, we further call it Misra fit, \cite{misra2020magnetic}) shows the best fit and the most accurate estimate of droplet parameters (droplet length and width) and outline. 
An example of how these approximations fit an elongated droplet can be seen in Figure \ref{fig:Misra_fit}.
Surprisingly, we find that the Misra fit describes poorly experimentally observed elongated magnetic droplets.
It overestimates the length of the magnetic droplet, thus skewing the data.
In contrast, ellipse and rectangle fits seem to appropriately estimate the magnetic droplet's parameters and seem to fit the shape well enough.
However, it was found that rectangular is the best fit resulting in lowest deviations from the droplet shape when comparing quantitatively.
Here we have to note that from the rectangle fit parameters we use side lengths to define the major axis $a$ and the minor axis $b$, which allow us to return to an ellipse shape description.

With ellipse parameters known, we can fit the experimental data with the relations \eqref{sigma_mu_eq} and \eqref{tau_eq} to find the values of surface tension, magnetic permeability and viscosity.
This is done using the least squares method.

\section{Results}
\label{sec:results}
\subsection{Characteristics for one sample and their dependence on time}
\begin{figure}
    \centering
    \includegraphics{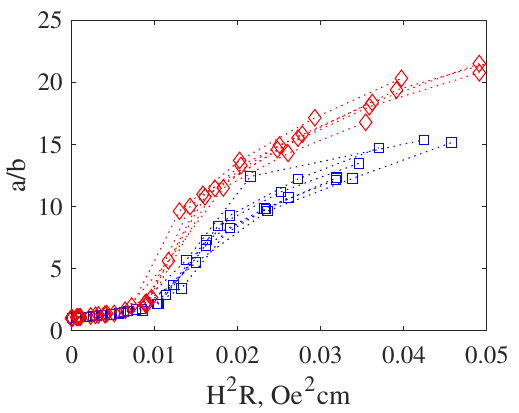}
    \caption{Data and corresponding fits of droplet deformation $a/b$ as a function of $H^2R$. Measurements were taken at $18^o$ C (red) and $8^o$C  (blue).}
    \label{fig:ab_curve}
\end{figure}

In order to obtain reproducible results, deformations of more than 200 droplets have been recorded and analyzed.
Fig.~\ref{fig:ab_curve} shows droplet shape deformations $a/b$ during the elongation experiment for multiple droplets at two different temperatures.
It is clearly visible that the curves are collected in two different sets, visually confirming the influence of temperature on the droplet characteristics.
However, it is of interest to examine the characteristics across a larger set of measurements.

\begin{figure}
    \centering
    \includegraphics[width=\textwidth]{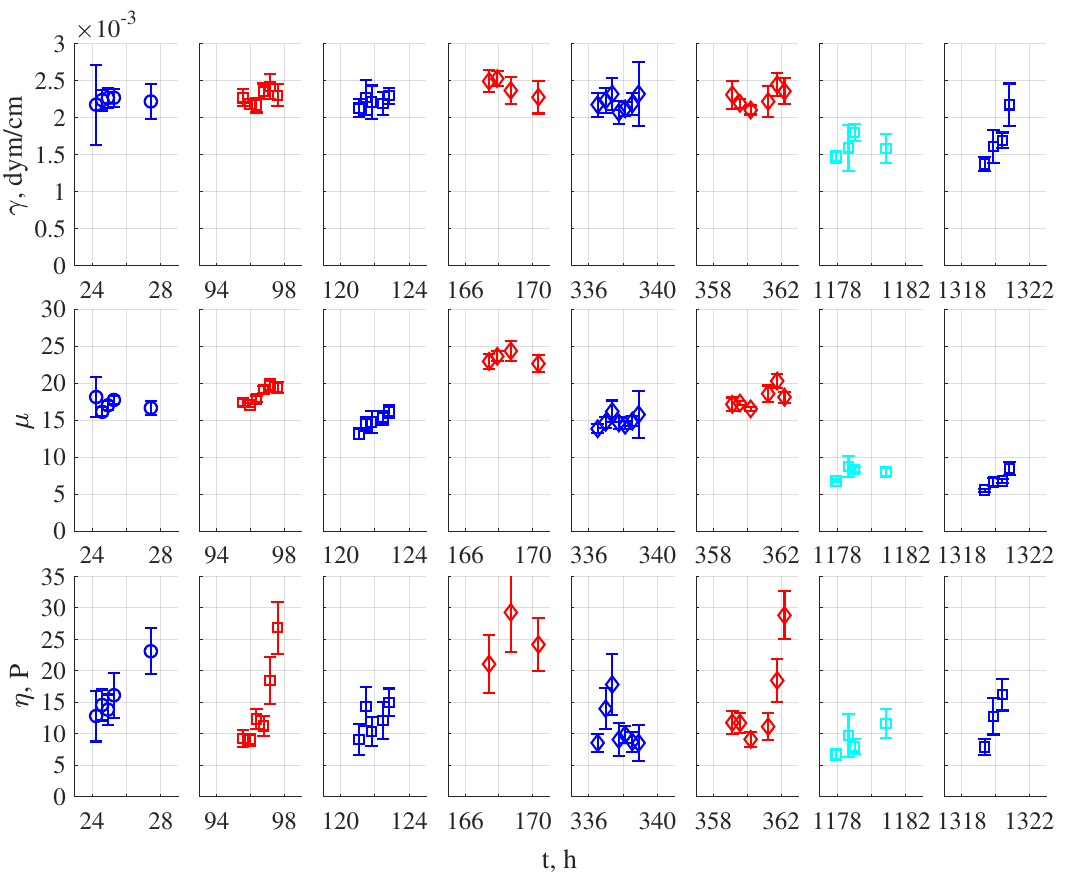}
    \caption{Surface tension $\gamma$ in dynes per cm (top row), magnetic permeability $\mu$ (dimensionless) (middle row) and viscosity $\eta$ in poise (bottom row) dependence on time, as measured by droplet elongation-relaxation experiments. Color represents temperature, while markers are the same for each of the samples.}
    \label{fig:sigma_time}
\end{figure}

Fitting models from sec.~\ref{sec:theory} to the experimental data allows us to obtain surface tension, magnetic permeability, and viscosity parameters for single droplets.
We consider that these measurements characterize the concentrated phase, assuming two simplifications about the dilute phase - (i) it can be considered non-magnetic and (ii) the viscosity is equal to that of water.
A summary of parameters determined for many droplets from a larger common destabilized sample ($\approx 1$~ml in volume) at three different temperatures is given in Fig.~\ref{fig:sigma_time}.
The different colors represent different temperatures.
The blue color is used for the data points at $8^{\circ}$C, red for $18^{\circ}$C and cyan for $13^{\circ}$C.
As measurements from multiple droplets in the same series give similar results, our assumption on using droplet measurement data for the characterization of the concentrated phase appears to be adequate.
However, we can see that there are several trends.

In general, there is a change in parameter values for different temperatures and over time.
Time here marks the time that has passed from the moment when the common destabilized sample was made.
For each set of measurements, one or several small capillaries are made from this common sample.
The surface tension $\gamma$ (Fig.~\ref{fig:sigma_time} top row) appears to almost stay constant around the value $2.0...2.5\cdot 10^{-3}$~dyn/cm, with the exception of measurements after $1000$~h from fluid destabilization. 
A different situation can be seen for magnetic permeability $\mu$, which can be seen in Fig.~\ref{fig:sigma_time}~middle row. 
Here, both time and temperature dependences appear to be important. 
In particular, at lower temperatures, the magnetic permeability is lower than at higher temperatures for consecutive series. 
Here again, measurements after $1000$~h give a significantly different picture, potentially indicating a stronger time dependence.  
Finally, viscosity, shown in Fig.~\ref{fig:sigma_time}~bottom row, shows a larger distribution in the measured values, including what appear to be outlayers, but again a tiny increase in viscosity can be seen with an increase in temperature. 
It has been difficult to obtain a higher quality of measurements, because during relaxation drops are much more sensitive to surface effects than during stepwise field alternation, where an equilibrium shape is sought.

The fourth column of data ($t\approx 167..171$~h) in Fig.~\ref{fig:sigma_time} were collected very soon after the small capillary sample was made, which could indicate that the liquid inside was not yet stabilized, complicating the precise measurements.  
Overall, this indicates that the rate at which the characteristic values change might also change depending on time and other conditions, suggesting that the rate of change is decreasing over time. 

It is worth to mention that many measurements were made with other samples in total at 6 temperatures in the range between $8..35^{\circ}$C, but for those the time of preparation of the destabilized sample was not recorded, making it impossible to present them in the same framework.
However, they have the same qualitative behavior as described here, giving more confidence in the reported results.

\subsection{Magnetic droplet characteristics dependence on temperature}

\begin{figure}
    \centering
    \includegraphics[width=1\textwidth]{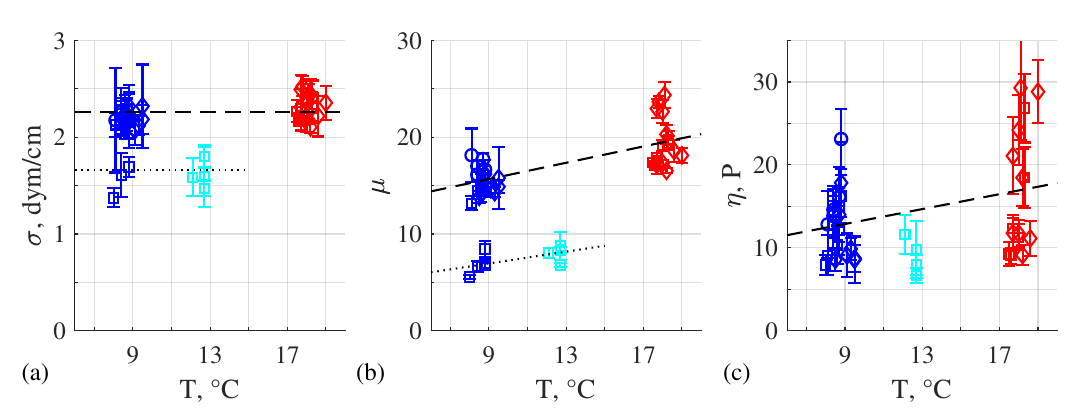}
    \caption{Surface tension $\gamma$ (a), magnetic permeability $\mu$ (b) and viscosity $\eta$ (c) dependence on temperature.}
    \label{fig:kopejie_rezultati}
\end{figure}

From the measurements shown previously, we can analyze them with respect to temperature $T$. 
This is done in Fig.~\ref{fig:kopejie_rezultati} for surface tension $\gamma$, magnetic permeability $\mu$ and viscosity $\eta$.
These two parameters are important for predicting the response of magnetic droplets and the concentrated phase in general.
As the surface tension coefficient appears to stay around the same value, we can consider that the temperature change affects only the magnetic permeability; as temperature increases, so does the magnetic permeability value.
This allows us to propose a temperature control mechanism and test it.

\begin{figure}
    \centering
    \includegraphics[width=1\textwidth]{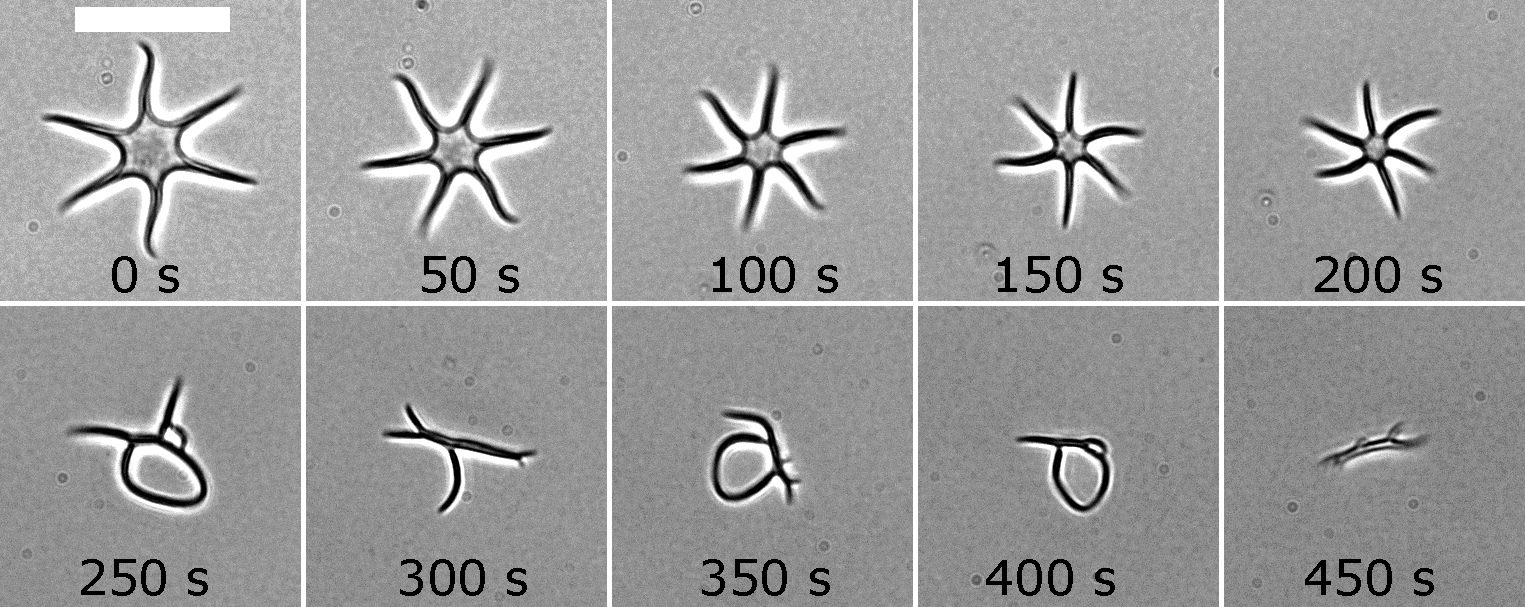}
    \caption{Reducing temperature during experiment causes the magnetic droplet to cross a critical field of the shape instability, transforming from a starfish to a prolate like object. Scale bar in top-right image is 25~$\mu$m}
    \label{fig:tuning}
\end{figure}
When exposed to rotating magnetic field, magnetic droplets are known to have a peculiar shape dynamics and undergo a re-entrant transition, that is, with an increase of magnetic field (or Bond number, to speak more generally), droplet first becomes an elongated prolate and with a further increase becomes flat oblate with spikes on the border, resembling a starfish \cite{dropPRL,erdmanis2017magnetic}.
These transitions are not only field or Bond number dependent, but also depend on the magnetic permeability $\mu$.
If $\mu$ is lower, a stronger field (or larger Bond number) is required for the transition to happen.

We tested our temperature control mechanism on the reentrant transition, and the result is shown in Fig.~\ref{fig:tuning} and Supplementary Video S1.
Initially, the parameters are set (rotating field with $H=13$~Oe, $f=50$~Hz, $T=22^{\circ}$C) to form a starfish-like droplet shape deformation.
Then we start to slowly decrease the temperature, while other parameters remain unchanged.
As expected, the starfish shape becomes unstable and gradually transforms into a prolate shape.
The process is slow due to the low surface tension. 
The droplet rotates with the field due to finite relaxation times.
However, we can see that also the size of the droplet has decreased, hinting that the temperature change influences not only $\mu$, but also the proportion of phase-separated material in the sample. 
Nevertheless, we have demonstrated the ability to use temperature as a tuning parameter. 

\section{Discussion}
\label{sec:discussion}
There are several aspects in our findings to compare to related works.
The first step is to choose the suitable form of the droplet shape approximation, which is at the core of all further analysis.
There has been notable criticism of the use of the ellipsoidal approximation \cite{Dropsimulation,misra2020magnetic}, argued by the inhomogenity of the inner magnetic field of the drop and justified by numerical simulations.
However, the only experimental comparison is with a single droplet that has a much smaller magnetic permeability $\mu$, smaller deformations but a much larger surface tension, which is considered field dependent, and larger field values, which go well into the nonlinear magnetization region \cite{pdmsdrops}.
Our experimental results show a slightly different picture. 
Even for large $\mu$ and large deformations $a/b$, the droplet shape resembles an ellipsoidal shape, although it is true that for $\mu>21$ this shape starts to deviate.
We find that a possible approximation is by using side lengths of a rectangle fit as more appropriate ellipse parameters. 
Moreover, the shape approximation proposed by Misra \cite{misra2020magnetic} does not work with the experimental data presented here.

During the early phases of magnetic fluid research it was theoretically predicted that phase separation can be induced in magnetic fluids if specific conditions are met \cite{victor1984liquid}.
Later in the literature it was shown that adding more salt has the same effect on the phase diagram as decreasing the temperature \cite{langmuir_og}, however little is known about the physical properties, including magnetic properties, surface tension and viscosity, and their change across the two phase coexistence curve.
Our results give some clues.
The first is that values of surface tension seem to stay constant, leaving place for interpretation in relation to phase-separation.
The second is that the magnetic permeability of the concentrated phase increases slightly with temperature.
Previous results suggest that the volume fraction $\Phi_c$ in the concentrated phase remains constant \cite{langmuir_og} and suggest that the relation between the volume fraction and the magnetic properties is not linear \cite{Ivanov}.

In this regard it is worth comparing these results with recent studies on magnetic fluid parameters in a phase-separated field-induced system \cite{Ivanov2018,Ivanov2019}.
There, an anomalous surface tension was observed, which usually increases with temperature. 
However, for lower temperatures and some samples, the surface tension values appear to remain constant in temperature ranges similar to those of this study.
It is interesting that the anomaly of the increase in surface tension is partially explained by phase-separation pecularities due to sample polydispersity, which means that larger particles tend to remain phase-separated for higher temperatures.
This principle is similar to the basis for the size classification used for magnetic fluids \cite{sizesort}. 
This idea offers an explanation for our observations: at higher temperatures, the average size of the magnetic nanoparticles in the phase-separated droplets increases, which means an increase in magnetic permeability due to larger magnetic moments for each of the particles and viscosity due to stronger magnetic interactions.
However, this needs to be confirmed.
Due to the very different magnetic fields that have been used in the studies, it is hard to make a more precise comparison with previous studies.

It is also worth noting some experimental limitations.
Magnetic droplets in the capillary sediment reach the lower surface of the capillary, as the high $\Phi_c$ gives a significantly denser fluid.
This means that there is a possibility for surface effects, including the physical severing of a droplet on an imperfection or rubble.
This can potentially interfere with decay measurements, by overestimating the relaxation time and viscosity values.
This might explain the overestimates in Fig.~\ref{fig:sigma_time}~(c). 
Additionally, a small extra magnetic field was used to compensate for parasitic magnetic fields and stabilize the magnetic droplets, making both droplet tips appear in focus.
Another factor to consider is the change in the properties of the sample over time.
After careful experiments, we show that there is a trend in the change of values depending on time since the magnetic fluid has been destabilized and when the magnetic fluid liquid is put into the capillary, indicating potential metastability.
In our earlier experiments high magnetic permeability $\mu$ values were found which complicated data processing, as mathematical models do not exist under those conditions. Because of this, a larger analysis was performed with magnetic droplets that have lower values of $\mu$. 

Our demonstration of tuning properties of phase-separated magnetic fluid opens up further research possibilities towards applications.
However, an important prerequisite would be to decorrelate the temperature effect on magnetic permeability and the volume of concentrated phase, as the latter changes the Bond number as a result of the change in the droplet radius.
A way to investigate this would be to use numerical simulations of magnetic droplet dynamics \cite{langins} with different properties.

\section{Conclusions}
\label{sec:coclusions}

The effect of temperature on the properties of a phase-separated magnetic fluid destabilized by an increase in ionic strength has been studied.
Analysis of droplet deformations under small fields at different temperatures revealed that the surface tension remains constant, whereas the magnetic permeability and viscosity of the concentrated phase slightly increase with temperature.
However, the investigated phase-separated magnetic fluid appears to change properties over longer times, reducing the precision of the measurements.
We also demonstrate a temperature-induced magnetic droplet shape transition, indicating possible applications.

\section{Acknowledgments}
The authors thank Prof. Andrejs Cebers for helpful advice and discussions.
We also thank the PHENIX lab of Sorbonne University and Dr. Emmanuelle Dubois and Prof. R\'{e}gine Perzynski in particular for the magnetic fluid and discussions.
This research is funded by the Latvian Council of Science, project BIMs, project No. lzp-2020/1-0149.

 \bibliographystyle{elsarticle-num} 
 \bibliography{bibliography}

\end{document}